\documentclass[twocolumn,showpacs,preprintnumbers,amsmath,prd,amssymb,superscriptaddress,chaproman]{revtex4}

\usepackage{graphicx}% Include figure files
\usepackage{dcolumn}% Align table columns on decimal point
\usepackage{bm}% bold math
\usepackage{multirow}
\usepackage{color}%

\usepackage[english]{babel}
\usepackage{amsmath,amssymb}
\usepackage{amsfonts}
\usepackage{stmaryrd}
\usepackage{mathrsfs}
\usepackage{epsfig}

\def\nuebar{{\rm \bar{\nu}_e}}
\def\nue{{\rm \nu_e}}
\def\s2tw{{\rm sin ^2 \theta_{W}}}
\def\Lambdanc{\Lambda_{NC}}

\begin{document}

%%\preprint{AS-TEXONO/12-01}

\title{Constraints on Non-Commutative Physics Scale
with Neutrino-Electron Scattering}

\newcommand{\as}{Institute of Physics, Academia Sinica, Taipei 11529, Taiwan.}
\newcommand{\metu}{Department of Physics,
Middle East Technical University, Ankara 06531, Turkey.}
\newcommand{\ktu}{Department of Physics,
Karadeniz Technical University, Trabzon 61080, Turkey.}
\newcommand{\deu}{Department of Physics,
Dokuz Eyl\"{u}l University, Buca, \.{I}zmir 35160, Turkey.} 
\newcommand{\thu}{Department of Engineering Physics, 
Tsinghua University, Beijing 100084, China.}
\newcommand{\bhu}{Department of Physics, Banaras Hindu University,
Varanasi 221005, India.}
\newcommand{\corr}{htwong@phys.sinica.edu.tw;
Tel:+886-2-2789-9682; FAX:+886-2-2788-9828.}

%%\affiliation{ \as }
%%\affiliation{ \metu }
%%\affiliation{ \ktu }
%%\affiliation{ \thu }
%%\affiliation{ \bhu }

\author{ S.~Bilmi\c{s}}  \affiliation{ \as } \affiliation{ \metu }
\author{ M.~Deniz }  \affiliation{ \as } \affiliation{ \ktu } \affiliation{ \deu}
\author{ H.B.~Li }  \affiliation{ \as }
\author{ J.~Li }   \affiliation{ \thu }
\author{ H.Y.~Liao }  \affiliation{ \as }
\author{ S.T.~Lin }  \affiliation{ \as }
\author{ V.~Singh }  \affiliation{ \as } \affiliation{ \bhu}
\author{ H.T.~Wong } \altaffiliation[Corresponding Author: ]{ \corr } \affiliation{ \as }
\author{ \.{I}.O.~Y{\i}ld{\i}r{\i}m}  \affiliation{ \as } \affiliation{ \metu }
\author{ Q.~Yue }  \affiliation{ \thu }
\author{ M.~Zeyrek } \affiliation{ \metu }

%%\collaboration{TEXONO Collaboration}  \noaffiliation

\date{\today}

\begin{abstract}

Neutrino-electron scatterings ($\nu - e$) are purely leptonic
processes with robust Standard Model (SM) predictions.
Their measurements
can therefore provide constraints to physics beyond SM.
Non-commutative (NC) field theories modify space-time
commutation relations, and allow neutrino electromagnetic
couplings at the tree level. 
Their contribution to neutrino-electron scattering
cross-section was derived. 
Constraints were placed on the 
NC scale parameter $\Lambdanc$ from 
$\nu - e$ experiments
with reactor and accelerator neutrinos.
The most stringent limit of 
$\Lambdanc > 3.3~{\rm TeV}$ at 95\% confidence level
improves over the direct bounds from 
collider experiments.

\end{abstract}

\medskip

%\keywords{
%%neutrinos in nonstandard model, 14.60.St
%%neutrinos interactions, 13.15.+g
%%Geometry noncommutative, 02.40.Gh
%%}

\maketitle

The physical origin and experimental consequences
of neutrino masses and mixings
are not fully understood or explored~\cite{pdg10numix}.
Experimental studies on the neutrino properties
and interactions 
can shed light to these
fundamental questions and
may provide hints or constraints to 
models on new physics.
Reactor neutrino is an excellent  
neutrino source to address many of
the issues, because of its high flux
and availability.
The reactor $\nuebar$ spectra is understood
and known, while reactor ON/OFF comparison
provides model-independent means of
background subtraction.

Neutrino-electron scatterings ($\nu - e$) are purely
leptonic processes with robust Standard Model (SM)
predictions~\cite{nuephys}.
It therefore provides an excellent probe
to physics beyond SM~\cite{nuephysbsm,texononsi}.
Experiments on $\nu - e$
scattering have played important
roles in testing SM, and
in the studies of neutrino intrinsic
properties and oscillation.
This article is a continuation of
our previous work in which
bounds were placed on
Non-Standard Interaction (NSI) parameters and
Unparticle physics~\cite{texononsi}.
The objective is to investigate
the consequences and constraints 
of non-commutative physics (NC) using
$\nu - e$ scattering data.

\begin{figure}
\begin{center}
\includegraphics[width=7cm]{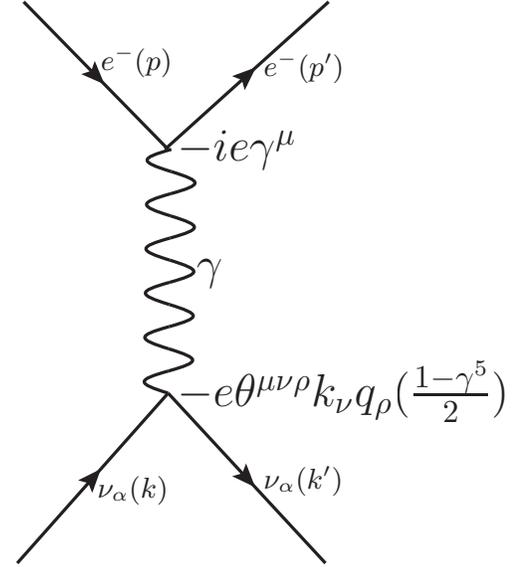} %%\\[2ex]
\end{center}
\caption{ \label{feydiagnc}
Schematic diagram of $\nu - e$ scattering via
virtual photon exchange in non-commutative space-time, 
where $\alpha = e, \mu ,\tau$ denotes the flavor state,
while k and p ($k^\prime$ and $p^\prime$) 
are the initial (final) four-momenta of 
the neutrino probe and electron target, respectively.
The NC coupling is given in Eq.~\ref{eq::ncem} while
$\theta^{\mu \nu \rho}$ is defined in 
Eq.~\ref{eq::theta}.
}
\label{fig::ncfeynman}
\end{figure}

\begin{table*} [hbt]
\caption{
Summary of experimental constraints on the NC energy scale 
$\Lambdanc$.  The quoted bounds for the direct experiments 
on scattering processes at colliders are at 95\% CL. 
These are complemented by order-of-magnitude estimates 
for the model-dependent bounds
with the atomic, hadronic and astrophysical systems.
The projected sensitivities from current and future
collider experiments are also listed.
}
\begin{ruledtabular}
\begin{tabular}{llc}
Experiments & Direct Scattering Channels & $\Lambdanc$ \\ \hline
\\
\multicolumn{3}{l}{High Energy Collider Experiments}\\
\multicolumn{3}{l}{~~ \underline{Current Bounds}}\\
~~~~~LEP-OPAL & 
$e^- + e^+ \rightarrow \gamma + \gamma $ \cite{opal} 
& ${\rm > 141~GeV}$ \\
~~~~~LEP & 
$e^- + e^+ \rightarrow Z \rightarrow \gamma + \gamma $ \cite{z_decay_bound} 
& $> 110$ GeV \\
~~~~~Tevatron &  $t \rightarrow W + b$ \cite{topquark_decay_width}
& $> 624$ GeV \\
& $t \rightarrow W_R + b$ \cite{topquark_decay_width}
& $> 1.5$ TeV \\
\multicolumn{3}{l}{~~ \underline{Projected Sensitivities}}\\
~~~~~LHC  & $Z \rightarrow \gamma + \gamma $ \cite{z_decay_width}
& $> 1$ TeV \\
& $p + p \rightarrow Z + \gamma \rightarrow l^+ + l^- + \gamma$ \cite{pp}
& $> 1$ TeV \\
 & $p + p \rightarrow W^+ + W^- $ \cite{ww}
& $> 840$ GeV \\
~~~~~Linear Collider  
& $e+\gamma \rightarrow e+\gamma$ \cite{compton}
& $> 900$ GeV \\
 & $e^- + e^- \rightarrow e^- + e^-$ \cite{bhabha} 
& $> 1.7$ TeV  \\
& $e^- + e^+ \rightarrow \gamma + \gamma$ \cite{bhabha}  
& $> 740$ GeV \\
& $\gamma + \gamma \rightarrow \gamma + \gamma$ \cite{bhabha} 
& $> 700$ GeV \\
& $e^- + e^+ \rightarrow \gamma + \gamma \rightarrow Z $ 
\cite{nc_2gamma} & $> 4$ TeV\\
& $e^- + e^+ \rightarrow Z + \gamma \rightarrow e^+ + e^- + \gamma$ 
\cite{pp} & $> 6$ TeV  \\
& $e^- + e^+ \rightarrow W^+ + W^- $ \cite{ww} 
& $> 10$ TeV  \\
~~~~~Photon Collider & 
$\gamma + \gamma \rightarrow l^+ + l^-$ \cite{photon}
& $> 700$ GeV \\
&$ \gamma + \gamma \rightarrow f + \bar{f}$ \cite{photon2}
& $> 1$ TeV \\
\\
\multicolumn{3}{l}{Low Energy and Precision Experiments }\\
\multicolumn{2}{l}{~~~~~Atom Spectrum of Helium \cite{helium} }
& $> 30$ GeV \\
\multicolumn{2}{l}{~~~~~Lamb Shift in Hydrogen \cite{hydrogen} }
& $> 10$ TeV \\
\multicolumn{2}{l}{~~~~~Electric Dipole Moment of Electron \cite{edm} }
& $> 100$ TeV \\
\multicolumn{2}{l}{~~~~~Atomic Clock Measurements \cite{clock} }
& $> 10^8$ TeV \\
\multicolumn{2}{l}{~~~~~CP Violating Effects in  $K^0$ System \cite{cp} } 
&  $> 2$ TeV \\
\multicolumn{2}{l}{~~~~~C Violating Effects in $\pi^0 \rightarrow 
\gamma + \gamma + \gamma$ \cite{pion_decay} }
& $> 1$ TeV \\
\multicolumn{2}{l}{~~~~~Magnetic Moment of Muon \cite{muon_magnetic_moment} }
& $> 1$ TeV \\
\\ 
\multicolumn{3}{l}{Astrophysics and Cosmology Bounds}\\
\multicolumn{2}{l}{~~~~~Energy Loss via $\gamma \rightarrow \nu \bar{\nu}$ 
in Stellar Clusters \cite{bound_astro} }
& $> 80$ GeV \\
\multicolumn{2}{l}{~~~~~Cooling of SN1987A 
via $\gamma \rightarrow \nu \bar{\nu}$ \cite{supernova} }
& $> 4$ TeV \\
\multicolumn{2}{l}{~~~~~Effects of $\gamma \rightarrow \nu \bar{\nu}$  
in Primordial Nucleosynthesis \cite{nucleosynthesis} }
& $> 3$ TeV \\
\multicolumn{2}{l}{~~~~~Ultra High 
Energy Astrophysical Neutrinos \cite{cosmic} }
& $> 200$ TeV \\
\\
\end{tabular}
\end{ruledtabular}
\label{tab::ncbounds}
\end{table*}

The differential cross-section in the rest frame of the initial electron for
$\nu_{\mu} ( \bar{\nu}_{\mu} ) -$e
elastic scattering in SM, where only neutral current is involved,
is given by~\cite{nuephys,texononue};
\begin{eqnarray} 
\left[ \frac{d\sigma ( ^{[-]} \hspace*{-0.35cm} {\nu}_{\mu} e ) }{dT} 
 ( E_{\nu} ) \right] _{SM}
& = &  \frac{G_{F}^{2}m_{e}}{2\pi }  \cdot
[ ~ \left(g_{V} \pm g_{A} \right) ^{2}  \nonumber \\
& + &  \left( g_{V} \mp g_{A} \right) ^{2}\left(1-
\frac{T}{E_{\nu }}\right) ^{2}  \nonumber  \\
& - & ( g_{V}^2 - g_{A}^2 ) ~ \frac{m_{e}T}{E_{\nu}^{2}} ~ ] ~~~ ,
\label{eq_cs_numu}
\end{eqnarray}
where $G_F$ is the Fermi coupling constant,
$T$ is the kinetic energy of the recoil electron,
$E_{\nu }$ is the incident neutrino energy, $m_e$ is mass of the electron 
and $g_{V}$, $g_{A}$ are the
vector and axial-vector couplings, respectively.
The upper(lower) sign refers to the
interactions with $\nu_{\mu} ( \bar{\nu}_{\mu} )$.
The coupling constants in SM can be expressed by
$g_{V}=-\frac{1}{2}+2\sin ^{2}\theta _{W}$
and $g_{A}=-\frac{1}{2}\label{eq_gvga}$, 
where $\s2tw$ is the weak mixing angle.
In $\nu_{e} ( \nuebar ) - e$ scattering,
all of charged-currents, neutral-currents and their
interference effects are involved~\cite{kayser79},
and the cross-section 
can be described through the replacement
$g_{V,A} \rightarrow ( g_{V,A} + 1 )$ in Equation~\ref{eq_cs_numu}.
Deviations of the measured electron recoil 
spectra with respect to
SM predictions would indicate new physics. 

Non-commutative (NC) field theories modify 
the space-time commutation relations. 
The idea dates back to the 1940's when it was used
to get rid of the divergences in quantum field theory 
before the renormalization concept was introduced~\cite{nc_history}. 
Recent revival of interest on NC physics 
comes with the study of NC space-time 
in string theories, quantum gravity
and Lorentz violation~\cite{nc_string,douglas,quantum_gravity,lorentz}.

The space-time coordinates in NC are considered as operators, 
with the commutation relation:
\begin{equation}
 [~ \hat{x}_\mu , ~ \hat{x}_\nu ~ ] ~ = ~ i ~ \theta_{\mu\nu} ~~ ,
\label{eq::ncom}
\end{equation}
where $\hat{x}_\mu$ denotes NC space-time coordinates. 
The real, antisymmetric matrix $\theta_{\mu\nu}$ 
is a constant with dimension 
${\rm (length)^2} \sim \rm{ (mass)^{-2}}$,
and represents the smallest 
observable area in the $( \mu , \nu )$ plane, 
analogous to the Planck constant
in space-momentum commutation relation. 
Ordinary space time relations are recovered 
at $\theta_{\mu\nu} = 0$. 

\begin{figure}[ht!]
\begin{center}
{\bf (a)}\\
\includegraphics[width=8.5cm]{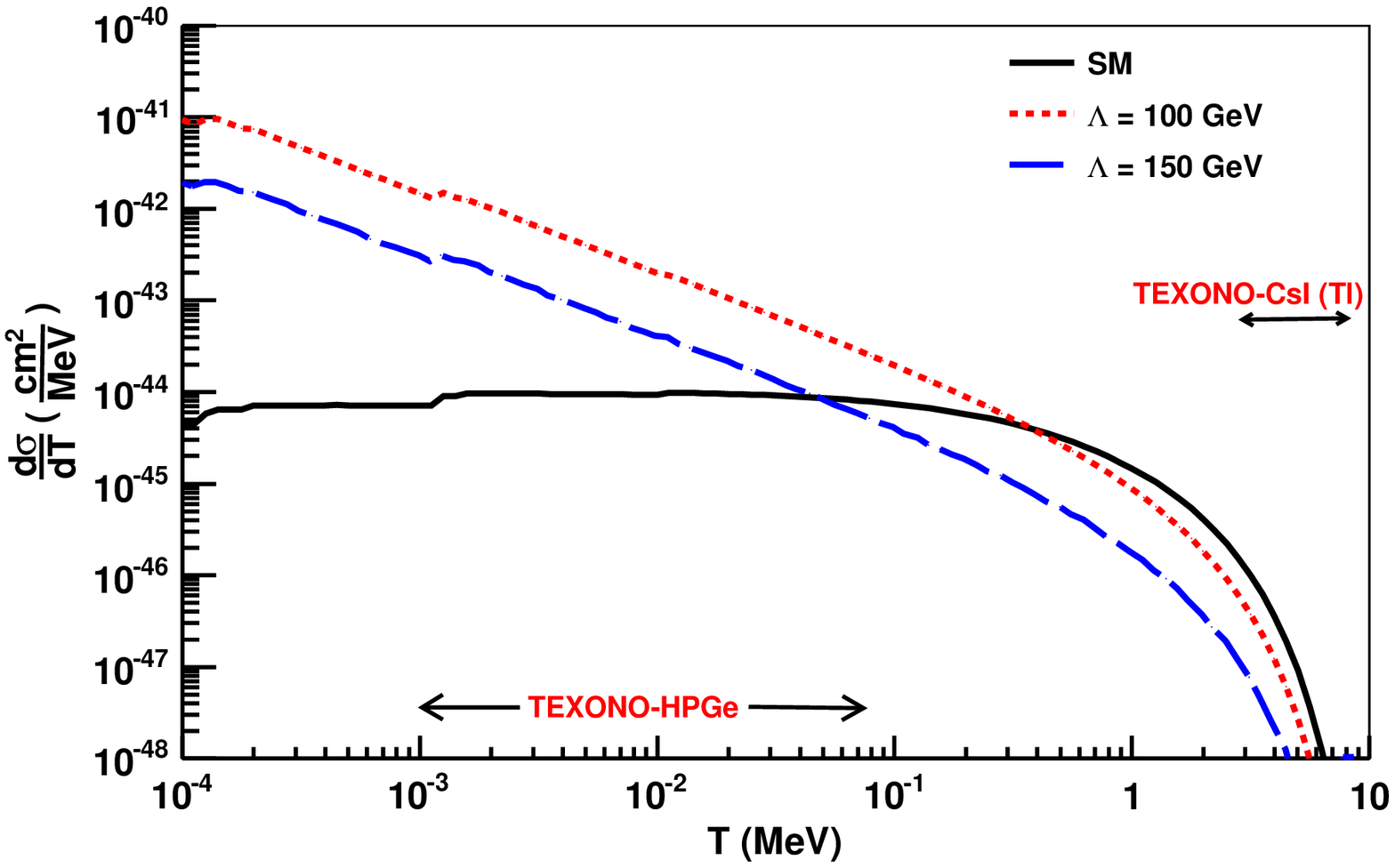} \\[3ex]
{\bf (b)} \\
\includegraphics[width=8.5cm]{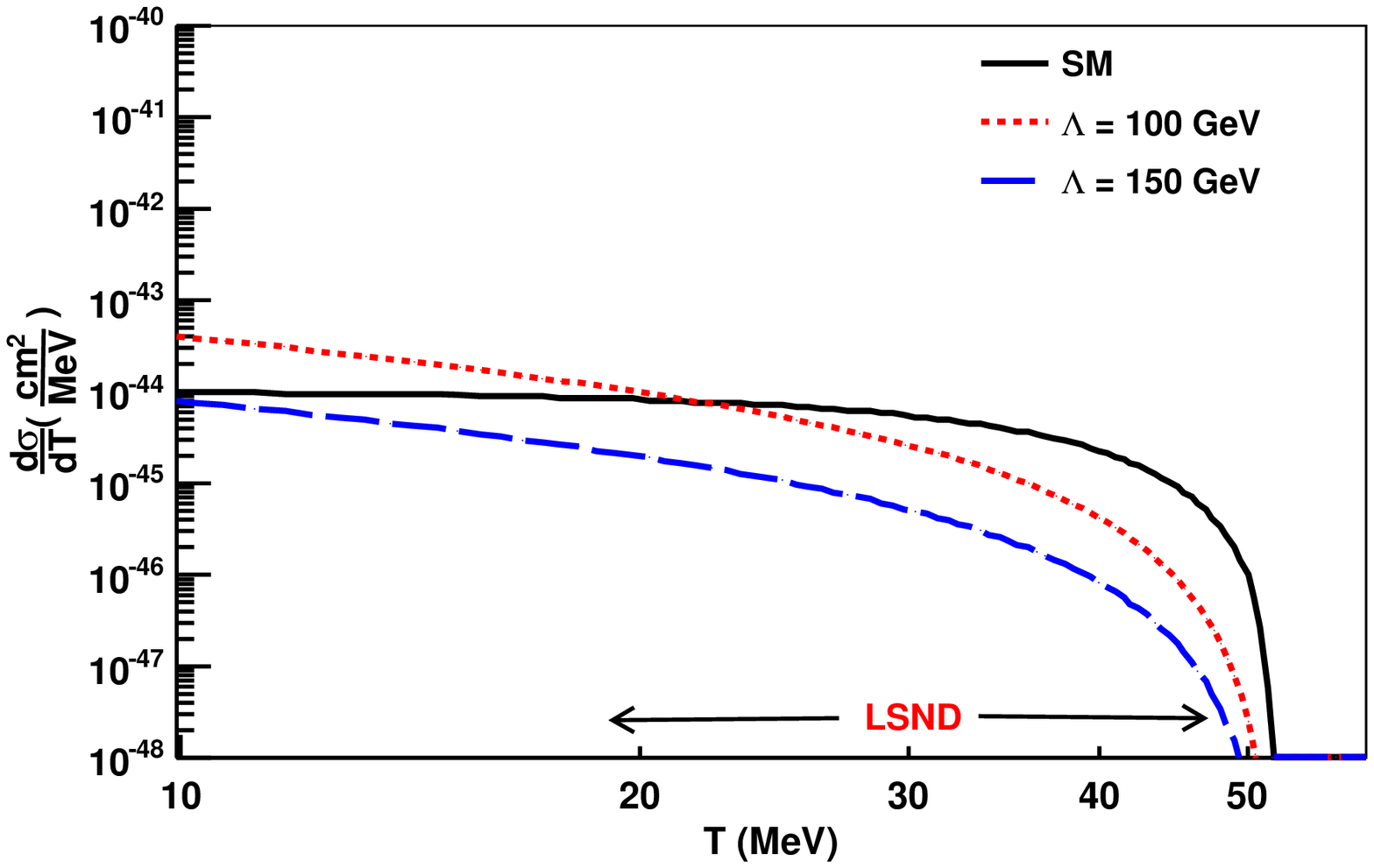} \\[2ex]
{\bf (c)} \\
\includegraphics[width=8.5cm]{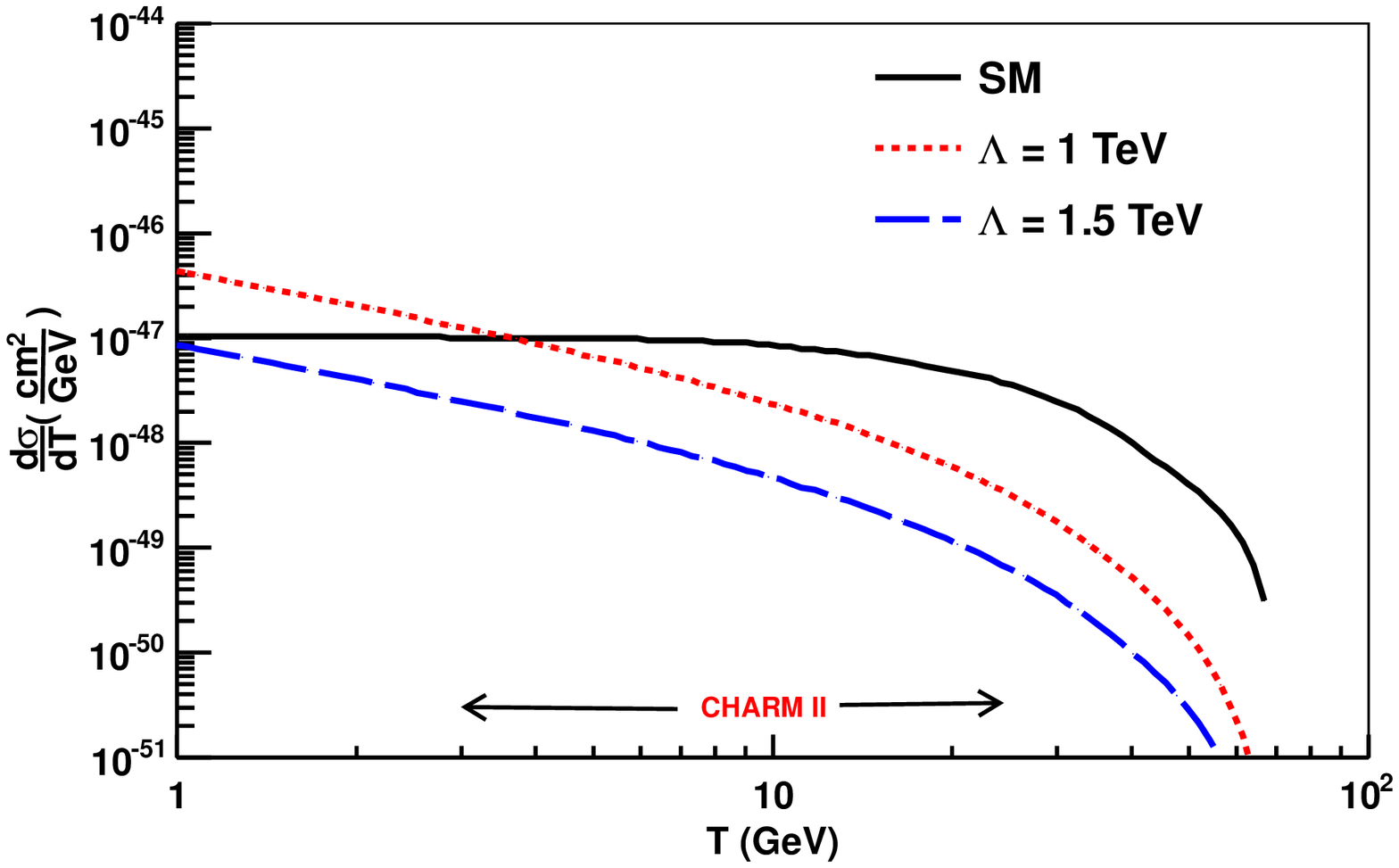} \\%%[2ex]
\end{center}
\caption{Differential cross-sections averaged over the
neutrino spectra as a function of the recoil energy
for (a) Top: TEXONO experiment
with reactor $\nuebar$~\cite{texonomunu,texononue},
(b) Middle: LSND experiment with $\nue$ from 
stopped-pion~\cite{lsnd}, and
(c) Bottom: CHARM-II experiment with accelerator 
$\nu_\mu(\bar{\nu}_{\mu})$~\cite{charm2}.
Both SM and NC contributions are displayed.
}
\label{fig::dsdT}
\end{figure}

Space-momentum commutation relation
gives rise to Heisenberg uncertainty principle.
Similarly, the commutation relation of Eq.~\ref{eq::ncom}
implies an uncertainty relation for space-time coordinates:
\begin{equation}
 \Delta x_\mu ~ \Delta x_\nu ~ \geq ~ \frac{1}{2} ~ |\theta_{\mu\nu}| ~ .
\end{equation}
NC physics is characterized by an
energy scale parameter
$\Lambdanc = ( 1 / \sqrt{|\theta_{\mu\nu}|} )$, below which
the space and time coordinates are incoherent.
New physics induced by NC effects would be important
above $\Lambdanc$. 
There is no theoretical bound on $\Lambdanc$.
Observable NC effects may be studied in
present and future experiments if $\Lambdanc$ would
be of the order of TeV.

Non-commutative 
field theory based on the commutation relation of 
Eq.~\ref{eq::ncom} has been constructed
via Weyl-Moyal star product~\cite{douglas}. 
The ordinary functions in Minkowski space time 
is retained via the definition of the star product 
which characterizes the NC structure. 
That is,
\begin{eqnarray}
 f(\hat{x}) g(\hat{x}) & \rightarrow & \hat{f}(x) * \hat{g}(x) \\ \nonumber
& = &
{\rm exp}~(\frac{i}{2} ~ \theta_{\mu\nu} ~ \partial_{x}^{\mu} ~ \partial_{y}^\nu )
~f(x)~g(y) ~ \mid _{x=y} ~~ .
\end{eqnarray}
The $\hat{~~}$ notation characterizes 
the NC-related variables which replace the SM analogs.
The lepton spinor and gauge field in NC space can be
expanded using Seiberg-Witten maps up to the 
first order of $\theta$ 
as~\cite{bound_astro,ettefaghi,ncsmnugamma}, respectively:
$$\hat\psi = \psi ~ + ~ e~ \theta^{\nu\rho} ~ A_{\rho} 
\partial_\nu \psi$$ 
and 
$$\hat{A_\mu} = A_\mu + e ~ \theta^{\nu\rho} ~ A_\rho ~
[\partial_\nu A_\mu - \frac{1}{2}\partial_\mu A_\nu] ~ .$$
Second order contributions in $\theta$ 
have also been worked out~\cite{theta2nd}.

Non-commutative QED has been 
constructed~\cite{bound_astro,supernova,Melic:2005fm}. 
Weyl-Moyal correspondence allows neutral
particles to couple with the $U(1)$ gauge field,
producing rich phenomenologies~\cite{ncsmnugamma,nc_2gamma}.
The action describing 
a neutral fermion field in NC-QED can be written
in terms of the usual field as~\cite{bound_astro}
\begin{eqnarray}
 S & = & \int d^4x
\bar{\psi}\big[(i\gamma^\mu\partial_\mu-m)  \\ \nonumber
&-& \frac{e}{2}\theta^{\nu\rho}(i\gamma^\mu(F_{\nu\rho}\partial_\mu+F_{\mu\nu}
\partial_\rho+F_{\rho\mu}\partial_\nu)-mF_{\nu\rho})\big]\psi  ~~ ,
\end{eqnarray}
where $F_{\mu\nu}= \partial_\mu A_\nu - \partial_\nu A_\mu$. 
Non-commutative Standard Model (NC-SM) 
has been constructed using analogous approach~\cite{ncsm}.

\begin{table*} [hbt]
\caption{The key parameters of
the TEXONO, LSND and CHARM-II measurements 
on $\nu - e$ scattering, and the derived bounds on NC physics.
The best-fit values in $\Theta ^2$ and 
the 95\% CL lower limits on 
$\Lambda_{NC}$ are shown.}
\begin{ruledtabular} 
\begin{tabular}{lcccccc}
Experiment  &  
$\nu$ & $< E_{\nu} >$ & $T$ & Measured $\s2tw$ &
Best-Fit on $\Theta ^2$ ($\rm{MeV^{-4}}$) 
& $\Lambda_{NC}$~(95\% CL) \\ \hline \\
TEXONO-HPGe~\cite{texonomunu} &
$\nuebar$ & 1$-$2~MeV & 12$-$60~keV & $-$ &
$(9.27  \pm 6.65 ) \times 10^{-22}$ &  $> 145~{\rm GeV}$ \\
TEXONO-CsI(Tl)~\cite{texononue} & 
$\nuebar$ & 1$-$2~MeV & 3$-$8~MeV &0.251 $\pm$ 0.039 &
$( 0.81 \pm 5.74 ) \times 10^{-21}$ &  $> 95~{\rm GeV}$ \\
LSND~\cite{lsnd} & 
$\nue$ & 36~MeV &18$-$50~MeV & 0.248 $\pm$ 0.051&
$(0.38 \pm 2.06) \times 10^{-21}$  & $> 123~{\rm GeV}$ \\
CHARM-II~\cite{charm2} &  
$\nu_{\mu}$ &23.7~GeV &3-24~GeV &
\multirow{2}*{ \} 0.2324 $\pm$ 0.0083 } &
$(0.20 \pm 1.03) \times 10^{-26}$  &  $> 2.6~{\rm TeV}$  \\
& $\bar{\nu}_{\mu}$ &19.1~GeV &3-24~GeV  & &
$(-0.92 \pm 4.77) \times 10^{-27}$  &  $> 3.3~{\rm TeV}$  \\
\end{tabular}
\end{ruledtabular}
\label{tab::results}
\end{table*}

Phenomenological studies of NC space-time  
and experimental constraints on $\Lambdanc$ 
were reviewed in Ref.~\cite{review}.
They are summarized in Table~\ref{tab::ncbounds}.
High energy collider experiments
probe possible NC-induced anomalous couplings 
between photons and leptons directly at the
relevant scale $\Lambdanc$.
The most stringent collider bounds 
are (i) $\Lambdanc > 141 ~ {\rm GeV}$ 
from the LEP-OPAL experiment~\cite{opal} based on 
NC-QED induced $e^- + e^+ \rightarrow \gamma + \gamma $, 
and (ii) $\Lambdanc > 1.5 ~ {\rm TeV}$ 
from the Tevatron experiments
via NC-SM induced the W-boson polarization 
in top quark decays~\cite{topquark_decay_width}.

Future experiments at LHC and linear collider may probe
$\Lambdanc <  10 ~ {\rm TeV}$.
The other approaches study indirect 
manifestations of the NC-effects, which
typically involve modeling of 
the atomic, QCD-hadronic and
astrophysical systems. 
Another implicit assumption 
for the low energy experiments
is that the formulation remains valid
at an energy-momentum range
significantly lower than $\Lambdanc$.

Neutrino-photon interactions  
are forbidden at the tree level in SM
and can proceed only as loop corrections.
However, NC-QED allows 
neutrino-photon couplings at tree
level due to new couplings to the 
U(1) gauge field~\cite{bound_astro}. 
Consequently, $\nu - e$ interactions 
can take place via exchange of virtual photons 
as depicted in Figure~\ref{fig::ncfeynman}.
This new channel will contribute to the measureable
$\nu - e$ cross-section 
in addition to the SM charged- and neutral-currents. 

The NC-QED coupling at the photon-neutrino vertex
is given by~\cite{bound_astro,ettefaghi}
\begin{equation}
 \Gamma^\mu (\nu \nu \gamma) = 
-e \theta^{\mu\nu\rho}k_\nu q_\rho(\frac{1 - \gamma_{5}}{2}) ~~ ,
\label{eq::ncem}
\end{equation}
where 
\begin{equation}
  \theta^{\mu\nu\rho}=\theta^{\mu\nu}
\gamma^\rho+\theta^{\nu\rho}\gamma^\mu+\theta^{\rho\mu} \gamma^\nu
\label{eq::theta}
\end{equation}
while k and p ($k^\prime$ and $p^\prime$) 
are the initial (final) four-momenta of the neutrino and electron,
respectively, and $q = p^\prime - p$. 
In contrast, the NC-QED photon-charged lepton vertex
coupling~\cite{bhabha,Ohl:2010zf} is given by
\begin{equation}
 \Gamma^\mu ( l^{\pm} l^{\pm} \gamma) =  
i e \gamma^\mu 
\exp [ i ( \frac{ p^\nu ~ \theta_{\nu \rho} ~ p^{\prime\rho}}{2}) ] ~, 
\end{equation}
in which the NC-effects manifest as anomalous phases. 
This characteristic feature
also applies to the case for 
NC-QED three- and four-photon 
coupling vertices~\cite{bhabha}.

Using the NC-QED
$\nu \nu \gamma$ vertex factor of Eq.~\ref{eq::ncem},
the matrix element for $\nu - e$ scattering in 
leading order of $\theta_{\mu\nu}$ 
can be written as: 
\begin{eqnarray}
 -i M_{NC} & = & \frac{e^2} {2 q^2} ~
\big[\bar{u}(p^\prime) \gamma^\mu u(p)\big]  \\ \nonumber
& & \big[\bar{u}(k^\prime)
\theta_{\mu\nu\rho}k^{\nu} q^\rho (\frac{1-\gamma_5}{2}) u(k)\big] ~.
\end{eqnarray}

Ignoring small neutrino mass, 
averaging over initial spin states 
and summing over final states,
the squared amplitude of NC contribution is
\begin{eqnarray}
 |M|^2 & = & \frac{32 e^4}{q^4} 
(\frac{\vec{\theta}\cdot\big(\vec{k}\times
\vec{k^\prime})}{2}\big)^2 \\ \nonumber
& & \big[(p \cdot k)(p^\prime \cdot k^\prime) 
+ (p \cdot k^\prime)(p^\prime \cdot k)-
m_e^2(k \cdot k^\prime) \big]  ~~ ,
\end{eqnarray}
where 
$\theta_{\mu\nu}k^\mu k^{\prime\nu} = \vec{\theta}\cdot (\vec{k}\times
\vec{k^\prime})$ 
under the definition $\vec{\theta} \equiv
(\theta_{23},\theta_{31},\theta_{12})$. 
Unitarity and causality relations
require $\theta_{0i} = 0$.~\cite{unitary}.
In the electron rest frame, 
$\vec{\theta}$ can be parametrized as 
$ \vec{\theta} = (\Theta \sin \xi , 0 , \Theta \cos \xi)$,
where $\xi$ is a phase angle and 
$\Theta = |\theta_{\mu\nu}| = ( 1 / \Lambda_{NC}^2)$.
Taking average over $\xi$, 
the NC-QED induced differential cross-section 
of $\nu - e$ scattering is:
\begin{eqnarray}
\label{eq::nc}
\left[ \frac{d\sigma (\nu  e)}{dT} ( E_{\nu} ) \right] _{NC} &=& 
\pi \alpha^2 \Theta^2 E_{\nu}^2  
~ \Big[ ~ \frac{1}{T} ~ - ~ \frac{2}{E_\nu} \\ \nonumber
& + & \frac{3T-2m_e}{2 E_{\nu}^2} ~ - ~ 
\frac{T^2 - 2 m_e T}{2 E_{\nu}^3} \\ \nonumber
& - & \frac{m_e T^2}{4 E_{\nu}^4} 
(1-\frac{m_e}{T}) ~ \Big] ~ .
\end{eqnarray}
The expression is 
valid for all types of $ \nu ( \bar{\nu} ) - e$
scatterings at $E_{\nu} \ll \Lambdanc$.
The $\alpha^2 \Theta^2 E_{\nu}^2$ dependence resembles 
that in NC-induced
$e^+ + e^- \rightarrow \nu_R + \bar{\nu}_R$~\cite{supernova}. 
There is no interference between the SM and NC channels in 
first order of $\Theta$~\cite{Melic:2005fm}. 
Interaction of $\nu - e$ via the exchange of Z and W-bosons 
with NC-SM is in principle possible.
However, in neutrino beam fixed-target experiments
where the four-momentum transfer is much smaller 
than Z-boson mass: $| q^2 | = 2 m_e T << m_Z^2$,
the NC-SM contribution is suppressed by 
$\sim [ ( m_e \cdot E_{\nu} ) / ( \alpha \cdot m_Z^2 ) ]^2$ 
relative to that of NC-QED in 
Eq.~\ref{eq::nc}, and hence can be neglected.

The $E_{\nu}^2$ and $1/T$ dependence in
the NC-QED cross-section of Eq.~\ref{eq::nc} is
significantly different from that of SM and
suggests that experimental investigations would
favor the studies of 
high energy neutrinos and low energy electron recoils. 
Three experiments with different ranges of 
neutrino energies
were selected for the analysis:
(I) TEXONO experiment with reactor $\nuebar - e$ at MeV range 
with $T \sim  10 - 100~{\rm keV}$ using high-purity
germanium detector~\cite{texonomunu}
and $ \sim 3 - 8~{\rm MeV}$ using CsI(Tl) crystal
scintillators~\cite{texononue};
(II) LSND experiment with $\nue - e$ in a stopped pion beam,
at  $T \sim  18 - 50~{\rm MeV}$ using 
liquid scintillator~\cite{lsnd}; and 
(III) CHARM-II experiment with high energy 
$\nu_\mu (\bar{\nu}_{\mu}) - e$ 
in a proton-on-target neutrino beam 
at $T \sim  3 - 24~{\rm GeV}$~\cite{charm2}. 
The key experimental parameters are summarized
in Table~\ref{tab::results}.
The differential cross-section in 
electron recoil energy
is given by:
\begin{equation}
< \frac{d \sigma}{dT} > ~ = ~
\int ~ \left[ \frac{d \sigma}{dT} ( E_\nu ) \right]
~ \phi_{\nu} ( E_{\nu} ) ~ d E_{\nu} ~~ ,
\label{eq::avedsdT}
\end{equation}
where the neutrino spectrum  $\phi_{\nu} ( E_{\nu})$
is normalized with
$ \int \phi_{\nu} ( E_{\nu}) d E_{\nu} = 1 $.
The measureable electron recoil spectra 
for the cases of SM and NC
in these experiment
are displayed in Figures~\ref{fig::dsdT}a,b\&c.
Both SM and NC contributions at typical values
of $\Lambda$ are overlaid.
The sawtooth structures for 
$T\lesssim ~10 {\rm ~ keV}$ in Figure~\ref{fig::dsdT}a 
are from correction due to atomic binding~\cite{binding}.

The combined contributions of SM and NC 
\begin{equation}
 \left( ~ \left[\frac{d \sigma}{dT} \right]_{SM} ~ +  ~
\left[ \frac{d \sigma}{dT} \right]_{NC}  ~\right)
\end{equation}
are placed in the integrand of 
Eq.~\ref{eq::avedsdT} and compared with
experimental data,
using the current values of $\s2tw$
at the respective ${\rm Q^2}$.
Constraints on $\Theta ^2$ 
are derived with a minimum-$\chi ^2$ analysis,
from which lower bounds of $\Lambda_{NC}$
at 95\% confidence level (CL) were derived.
The results are summarized in Table~\ref{tab::results}.
With the sensitivities enhanced by a $E_{\nu}^2$ factor,
the most stringent lower limit comes from the 
CHARM-II experiment 
with high energy accelerator neutrinos, where 
\begin{equation}
\Lambda_{NC} > 3.3 ~ {\rm TeV} 
\end{equation}
at 95\% CL.
This improves over the best 
direct bounds from collider
experiments. 
We note also that a similar analysis was attempted 
with reactor neutrinos data~\cite{texonomunu}
in Ref.~\cite{ettefaghi}.
However, an error in the differential
cross-section formula 
equivalent to Eq.~\ref{eq::nc} 
%%(the last term was 
%%incorrectly stated as $[1 - (m_e/E_{\nu})]$)
together with a missing factor in $\alpha$ 
in the numerical evaluation 
make the results invalid.

It is possible that NC physics can also 
give rise to flavor-changing 
transitions in $\nu - e$ scattering. 
The NC analysis of this work can be extended to include 
two parameters ($\Lambdanc , \lambda_{\alpha \beta}$),
where $\lambda_{\alpha \beta}$ 
denotes the branching ratio 
of the flavor-changing process
$\nu_{\alpha} + e \rightarrow \nu_{\beta} + e$.
The bounds would be relevant to the analysis
of the precision neutrino oscillation measurements,
in which sub-leading NC-induced effects may appear 
at the sources, during propagation through matter 
and and at the detectors. Such studies would be 
analogous to the combined analysis 
of non-standard neutrino interactions 
and oscillation parameters~\cite{nuoscnsi}.

The authors appreciate discussions and comments from
Drs. T.M.~Aliev, 
M.M Ettefaghi, X.G.~He and
A.~\"{O}zpineci. 
This work is supported by 
Contract No. 108T502  under TUBITAK, Turkey
and 99-2112-M-001-017-MY3 
under the National Science Council, Taiwan.

\end{document}